# A No-Go Theorem About Rotation in Relativity Theory[1]

(To appear in a *Festschrift* for Howard Stein,
edited by David B. Malament, and published by Open Court Press.)

David B. Malament
Department of Logic and Philosophy of Science
The University of California, Irvine
3151 Social Science Plaza
Irvine, CA 92697-5100
dmalamen@uci.edu

Within the framework of general relativity, in some cases at least, it is a rather delicate and interesting question just what it means to say that a body is or is not "rotating". Moreover, the reasons for this -- at least the ones I have in mind -- do not have much to do with traditional controversy over "absolute vs. relative" conceptions of motion. Rather they concern particular geometric complexities that arise when one allows for the possibility of spacetime curvature. The relevant distinction for my purposes is not that between attributions of "relative" and "absolute" rotation, but between attributions of rotation than can and cannot be analyzed in terms of a motion (in the limit) at a point. It is the latter -- ones that make essential reference to extended regions of spacetime -- that can be problematic.

The problem has two parts. First, one can easily think of different criteria for when a body is rotating. The criteria agree if the background spacetime structure is sufficiently simple, e.g., in Minkowski spacetime (the regime of "special relativity"). But they do not do so in general. Second, none of the criteria fully answers to our classical intuitions. Each one exhibits some feature or other that violates those intuitions in a significant and interesting way.

My principle goal in what follows is to make the second claim precise in the form of a modest no-go theorem. To keep things simple, I'll limit attention to a special case. I'll consider (one-dimensional) rings centered about an axis of rotational symmetry, and consider what it could mean to say that the rings are *not* rotating around the axis. (It is convenient to work with the negative formulation.) The discussion will have several parts.

First, for purposes of motivation, I'll describe two standard criteria of non-rotation that seem particularly simple and natural. (I could assemble a longer list of proposed criteria, but I am more interested in formulating a general negative claim that applies to all.[2]) One involves considerations of angular momentum ("ZAM criterion"). The other is cast in terms of the "compass of inertia" on the axis ("CIA criterion"). Next, I'll characterize a large class of "generalized criteria of non-rotation" that includes the ZAM and CIA criteria. Third, I'll abstract two (seemingly) modest conditions of adequacy that one might expect a criterion of non-rotation to satisfy (the "limit condition" and the "relative rotation condition"). Finally, I'll show that no (non vacuous)[3] "generalized criterion of non-rotation" satisfies both conditions in all relativistic spacetime models. The proof of the theorem is entirely elementary once all the definitions are in place. But it may be of some interest to *put* them in place and formulate a result of this type. The idea is to step back from the details of particular proposed criteria of non-rotation and direct attention instead to the conditions they do and do not satisfy.

## I. Informal Preview

Beginning in section II, our discussion will be cast in the precise language of relativistic spacetime geometry. But first, to explain and motivate what is coming, we give a rough, preliminary description of the no-go theorem in more direct, intuitive, quasi-operational terms. This will involve a bit of hand-waving, but not much. (This section will not presuppose familiarity with the mathematical formalism of general relativity.)

Consider a ring positioned symmetrically about a central axis as in figure 1.[4]  At issue is what it means to say that the ring is *not-rotating* (about that axis). The first criterion we will be considering takes the absence of inertial or dynamical effects on the axis as the standard for non–rotation.  Here is one way to set things up in terms of a telescope and a water bucket.  (Water buckets, to be sure, are not particularly sensitive instruments, but they are good enough for our purposes.) Let P be the point of intersection of the axis with the plane of the ring.  Place a lazy susan at P (in the plane of the ring), bolt a half-filled water bucket to the center of the lazy susan, and bolt a tubular telescope to the water bucket.  (See figure 2.)  Finally, mount a light source at a point (any point) on the ring.  Now consider possible rotational states of the composite apparatus on the axis (lazy susan + water bucket + telescope). There is one state in which the apparatus *tracks the ring* in the sense that an observer, standing on the lazy susan and looking through the telescope, will see the light source permanently fixed on its cross hairs. We take the ring to be *non-rotating according to the CIA criterion* if in this state (the tracking state), the water surface in the bucket is flat rather than concave.

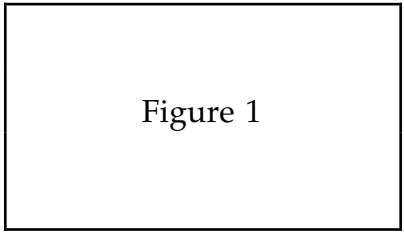

Figure 1

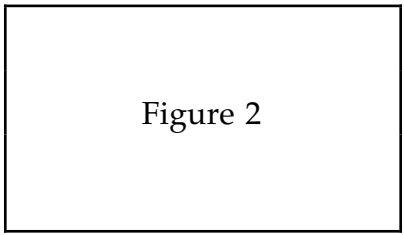

Figure 2

This characterization is a bit complicated because it makes use of a telescope as well as a water bucket. The former is used to bridge the distance between the water bucket here and the ring there.

We can actually use the instruments described to ascribe an angular speed to the ring (relative to the compass of inertia on the axis). Let the composite apparatus be placed in a state of motion in which the water surface *is* flat. And (just to keep things simple), let us assume that, at some initial moment, the observer standing on the lazy susan sees the light source through the telescope. It may be the case that he continues to see it as time elapses. (This is just the case in which the ring is judged to be non-rotating according to the CIA criterion.) But, in general -- assuming the ring is in *some* state of uniform rotational motion -- he will see it periodically, with a characteristic interval of time $\Delta t$ between sightings. (We imagine that the observer carries a stopwatch.) This interval is the time it takes for the ring to complete one rotation (relative to the CIA). So the angular speed of the ring (relative to the CIA) is just $2\pi/\Delta t$.

Now we consider a second criterion of non-rotation that is, on the face of it, very different in character from the first. There is a generic connection in mechanics, whether classical or relativistic, between (continuous) symmetries of spacetime structure and conserved quantities. Associated with the rotational symmetry of the ring system under consideration is a notion of angular momentum.[5] According to our second (ZAM) criterion, the ring is "non-rotating" precisely if the value of that angular momentum is zero at every point on the ring. The condition has an intuitive geometrical interpretation that we will review later. Here, instead, we describe an experimental test for determining whether the condition obtains. (Many other tests could be described just as well.)

Imagine that we mount a light source at some point Q on the ring, and from that point, at a given moment, emit light pulses in opposite (clockwise and counterclockwise) directions. This can be done, for example, using concave mirrors attached to the ring. Imagine further that we keep track of whether the pulses arrive back at Q simultaneously (using, for example, an interferometer). It turns out that this will be the case -- they will arrive back simultaneously -- if and only if the ring has zero angular momentum.

This equivalence is not difficult to verify and we will do so later. But wholly apart from the connection to angular momentum, the experimental condition described should seem like a natural criterion of non-rotation. Think about it. Suppose the ring is rotating in, say, a counterclockwise direction. (Here I am just appealing to our ordinary intuitions about "rotation".) The C pulse, the one that moves in a clockwise direction, will get back to Q before completing a full circuit of the ring because it is moving toward an approaching target. In contrast, the CC pulse is chasing a receding target. To get back to Q it will have to traverse the entire length of the ring, and then it will have to cover the distance that Q has moved in the interim time. One would expect, in this case, that the C pulse would arrive back at Q before the CC pulse. (Presumably light travels at the same speed in all directions.) Similarly, if the ring is rotating in a clockwise direction, one would expect that the CC pulse would arrive back at Q before the C pulse. Only if the ring is not rotating, should they arrive simultaneously. Thus, our experimental test for whether the ring has zero angular momentum provides what would seem to be a natural criterion of non-rotation. (Devices working on this principle, called "optical gyroscopes", are used in sensitive navigational systems. See, for example, the discussion in Ciufolini and Wheeler (1995), p. 365.)

We now have two criteria for whether the ring is non-rotating. It is non-rotating in the first sense if it is non-rotating with respect to the compass of inertia on the axis (as determined, say, using a water bucket and telescope). It is non-rotating in the second sense if it has zero angular momentum (as determined, say, using light pulses circumnavigating the ring in opposite directions). As we shall see later, it is a contingent matter in general relativity whether they agree or not. Whether they do so depends on the background spacetime structure in which the ring is imbedded. If it is imbedded in Minkowski spacetime, for example, it will qualify as non-rotating according to the CIA criterion iff it does so according to the ZAM criterion. But *if it is imbedded, instead, for example, in Kerr spacetime, the equivalence fails*. (We choose this example lest one imagine that the failure of agreement occurs only in pathological spacetime models that are of mathematical interest only. The Kerr solution may well describe regions of our universe, the real one, at least approximately -- regions surrounding rotating black holes.)

Though the two criteria do not agree in general, it is important for our purposes that *they "agree in the limit for infinitely small rings", no matter what the background spacetime structure*. They do so in the following sense. Imagine that we have a sequence of rings $R_1, R_2, R_3, ...$ that share a center point P on the axis, and have radii that shrink to 0. Imagine further that each of them is non-rotating according to the ZAM criterion. Each ring $R_i$ has a certain angular speed $\omega_i$ with respect to the compass of inertia on the axis. (We described a procedure above for measuring it.) None of the $\omega_i$ need be 0. The claim here is that (regardless of the background spacetime structure), the sequence $\omega_1, \omega_2, \omega_3, ...$ must converge to 0. (We will verify the claim later.)

We have considered just two simple, natural, experimental criteria for non-rotation. We could consider others (that do not, in general, agree with either one). But the fact is, it would turn out in every case that the criterion agrees with them

"in the limit for infinitely small rings" in the sense just described (no matter what the background spacetime structure).[6] This is one way to understand the claim that there *is* a robust notion of rotation (in the limit) at a point in general relativity, even if there is none that applies to extended regions of spacetime.

In any case, with these remarks as motivation, we now propose for consideration a first condition that one might expect a reasonable criterion of non-rotation to satisfy. Let us understand a "generalized criterion of non-rotation" to be, simply, a specification, for every ring, in every state of motion (or non-motion), whether it is to qualify as "non-rotating". We don't require that it have a natural geometrical or experimental interpretation.

> <u>Limit</u> <u>Condition</u>:   Let $R_1, R_2, R_3, ...$ be a sequence of rings, each "non-rotating", that share a center point P on the axis, and have radii that converge to 0. For every i, let $R_i$ have angular speed $\omega_i$ with respect to the compass of inertia on the axis. Then the sequence $\omega_1, \omega_2, \omega_3, ...$ converges to 0.

We have just asserted that the ZAM criterion satisfies this condition (regardless of the background spacetime structure). The CIA criterion does too, of course. (In the latter case, $\omega_i$ is 0, for every i. So the sequence certainly converges to 0.)

It remains to state our second condition (on a generalized criterion of non-rotation).  Suppose we have two rings $R_1$ and $R_2$ centered about the axis as in figure

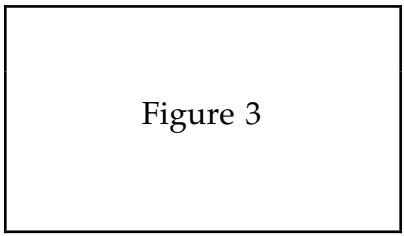

Figure 3

3.  (The planes of the rings are understood to be parallel, but nor necessarily

coincident.) Further suppose that "$R_2$ is non-rotating relative to $R_1$". Then, one would think, either both rings should qualify as "non-rotating", or neither should. This is precisely the requirement captured in our "relative rotation condition". It is not entirely unambiguous what it means to say that $R_2$ is non-rotating *relative* to $R_1$. But all we need is a sufficient condition for relative non-rotation of the rings. And it seems, at least, a plausible sufficient condition for this that, over time, there is no change in the distance between any point on one ring and any point on the other, i.e., the two rings move as if locked together.

> <u>Relative Rotation Condition</u>: Given two rings $R_1$ and $R_2$, if (i) $R_1$ is "non-rotating", and if (ii) $R_2$ is non-rotating relative to $R_1$ (in the sense that, given any point on $R_2$ and any point on $R_1$, the distance between them is constant over time), then $R_2$ is "non-rotating".

The relative rotation condition is really at the heart of our discussion. It seems a modest condition. But neither the CIA nor the ZAM criterion satisfies it, in general! They both do so if the rings are imbedded in Minkowski spacetime. But, as we shall see, *neither does if they are imbedded in, for example, Kerr spacetime.*

It should be clear just what is being asserted here. The situation is extremely counterintuitive. Consider the ZAM criterion. The claim is that we can have two rings, moving as if rigidly locked together, where one, but not the other, has zero angular momentum. Light pulses circumnavigating the first will arrive back at their starting point simultaneously. But pulses circumnavigating the second will not do so. (And similarly with the CIA criterion.)

We have made a number of claims involving two criteria of non-rotation, two possible conditions of adequacy on a criterion of rotation, and two spacetime models. It may help to summarize some of those claims in a table.

|  | In Minkowski spacetime | In Kerr spacetime |
|---|---|---|
| Do the CIA and ZAM criteria agree (for rings of arbitrary radius)? | Yes | No |
| Does the CIA criterion satisfy the limit condition? | Yes | Yes |
| Does the ZAM criterion satisfy the limit condition? | Yes | Yes |
| Does the CIA criterion satisfy the relative rotation condition? | Yes | No |
| Does the ZAM criterion satisfy the relative rotation condition? | Yes | No |

The fact that neither the CIA nor ZAM criterion satisfies the relative rotation condition (in general), seems a significant strike against them, and it is natural to ask whether any other criterion does better. Our principal claim is that, in an interesting sense, the answer is 'no'. There *are* criteria that satisfy the relative rotation condition in Kerr spacetime.[7] But the cost of doing so is violation of the limit condition, or else the radical conclusion that no ring in *any* state of motion (or non-motion) counts as "non-rotating".

<u>Theorem</u>  In Kerr spacetime (and other relativistic spacetime models to be discussed), there is no generalized criterion of non-rotation that satisfies the following three conditions:

(i)  limit condition

(ii)  relative rotation condition

(iii)  non-vacuity condition:  there is some ring in some state that qualifies as "non-rotating".

The result is intended to bear this interpretation. *Given any (non-vacuous)*

*generalized criterion of non-rotation in Kerr spacetime, to the extent that it gives "correct" attributions of non-rotation in the limit for infinitely small rings -- the domain where one does have a robust notion of non-rotation -- it must violate the relative rotation condition.*

## II. Formal Treatment

Now we start all over and cast our discussion in the language of relativistic spacetime time geometry.[8] We present formal versions of the two criteria of non-rotation and the two conditions of adequacy (though not in the same order as in section I).

First we have to consider how to represent one-dimensional rings in a state of uniform rotational motion. To keep things as simple as possible, we will think of the rings as test bodies with negligible mass, imbedded in a background spacetime structure that exhibits the rotational and "time translational" symmetries of the ring system itself. More precisely, we will think of them as imbedded in a stationary, axi-symmetric spacetime model.

### II.1 Stationary, Axi-Symmetric Spacetimes

We take a *(relativistic) spacetime model* to be a structure $(M, g_{ab})$ where M is a connected, smooth, four–dimensional manifold, and $g_{ab}$ is a smooth, pseudo–Riemannian metric on M of Lorentz signature $(+, -, -, -)$. We say that $(M, g_{ab})$ is *stationary and axi-symmetric* if there exist two one-parameter isometry groups acting on M, $\{\Gamma_t: t \in \mathbf{R}\}$ and $\{\Sigma_\varphi: \varphi \in \mathbf{S}^1\}$, satisfying several conditions. (Here we identify $\mathbf{S}^1$ with the set of real numbers mod $2\pi$.)[9]

(SAS 1) The isometries $\Gamma_t$ and $\Sigma_\varphi$ commute for all $t \in \mathbf{R}$ and $\varphi \in \mathbf{S}^1$.

(SAS 2) $\Gamma_t(p) \neq p$ for all points p in M and all $t \neq 0$. (So the orbits of all points under $\{\Gamma_t: t \in \mathbf{R}\}$ are open.)

(SAS 3) Some, but not all, points p in M have the property that $\Sigma_\varphi(p) = p$ for all $\varphi$. (Those with the property are called *axis points*. So the orbits of axis points under $\{\Sigma_\varphi: \varphi \in \mathbf{S}^1\}$ are singleton sets, and those of non-axis points are (non-degenerate) closed curves.)

(SAS 4) The orbits of $\{\Gamma_t: t \in \mathbf{R}\}$ are timelike, and the non-degenerate orbits of $\{\Sigma_\varphi: \varphi \in \mathbf{S}^1\}$ are spacelike.[10]

The final condition is slightly more complex than the others. Let $M^-$ be the restricted manifold that one gets by excising the (closed) set of axis points. The orbit of any point in $M^-$ under the two-parameter isometry group $\{\Gamma_t \circ \Sigma_\varphi: t \in \mathbf{R}\ \&\ \varphi \in \mathbf{S}^1\}$ is a smooth, two-dimensional, timelike[11] submanifold that is diffeomorphic to the cylinder $\mathbf{R} \times \mathbf{S}^1$. Let us call it an *orbit cylinder*. So (in the tangent space) at every point p in $M^-$, there is a timelike two-plane T(p) tangent to the orbit cylinder that passes through p, and a spacelike two-plane S(p) orthogonal to T(p). The final condition imposes the requirement that the set $\{S(p): p \in M^-\}$ be integrable.

(SAS 5) (Orthogonal transitivity) Through every point in $M^-$ there is a smooth, two-dimensional, spacelike submanifold $\Pi$ that is tangent to S(p) at every point p in $\Pi \cap M^-$.

Associated with the two isometry groups $\{\Gamma_t: t \in \mathbf{R}\}$ and $\{\Sigma_\varphi: \varphi \in \mathbf{S}^1\}$, respectively, are Killing fields $\tau^a$ and $\varphi^a$. (They arise as the "infinitesimal generators" of those groups.) It will be helpful for what follows to reformulate the five listed conditions in terms of these fields. (SAS 1) is equivalent to the assertion that the fields have vanishing Lie bracket, i.e., at all points

$$\tau^n \nabla_n \varphi^a - \varphi^n \nabla_n \tau^a = 0. \tag{1}$$

(SAS 2) comes out as the requirement that $\tau^a$ be everywhere non-zero. (SAS 3) can be understood to assert that $\varphi^a$ vanishes at some (axis) points, but does not vanish everywhere. (SAS 4) is equivalent to the assertion that $\tau^a$ is everywhere timelike, and $\varphi^a$ is spacelike at non–axis points. Finally, (SAS 5) is equivalent (by Frobenius' theorem) to the assertion that the conditions

$$\varphi_{[a}\tau_b \nabla_c \tau_{d]} = 0 \tag{2a}$$

$$\tau_{[a}\varphi_b \nabla_c \varphi_{d]} = 0 \tag{2b}$$

hold at every non-axis point.[12]

The five listed conditions imply the existence of coordinate functions with respect to which the metric $g_{ab}$ assumes a special, characteristic form.[13] The first three imply that there exist smooth maps $t: M \to \mathbf{R}$ and $\varphi: M^- \to \mathbf{S}^1$ such that $\tau^a = (\partial/\partial t)^a$ on M, and $\varphi^a = (\partial/\partial \varphi)^a$ on $M^-$. (Again, $M^-$ is the restricted submanifold on which $\varphi^a \neq 0$.) The remaining conditions imply that, at least locally on $M^-$, we can find further smooth coordinates $x_2$ and $x_3$ such that, at every point, the vectors $(\partial/\partial x_2)^a$ and $(\partial/\partial x_3)^a$ are spacelike, orthogonal to each other, and orthogonal to both $(\partial/\partial t)^a$ and $(\partial/\partial \varphi)^a$. Thus, at points in $M^-$, the matrix of components of $g_{ab}$ with respect to the coordinates $(t, \varphi, x_2, x_3)$ has the characteristic form:

| $\tau^a \tau_a$ | $\tau^a \varphi_a$ | 0 | 0 |
| $\tau^a \varphi_a$ | $\varphi^a \varphi_a$ | 0 | 0 |
| 0 | 0 | $(\partial/\partial x_2)^a (\partial/\partial x_2)_a$ | 0 |
| 0 | 0 | 0 | $(\partial/\partial x_3)^a (\partial/\partial x_3)_a$ |

And the inverse matrix (giving the components of $g^{ab}$) has the form:

$$\begin{array}{cccc}
(\varphi^a\varphi_a)D^{-1} & (-\tau^a\varphi_a)D^{-1} & 0 & 0 \\
(-\tau^a\varphi_a)D^{-1} & (\tau^a\tau_a)D^{-1} & 0 & 0 \\
0 & 0 & [(\partial/\partial x_2)^a(\partial/\partial x_2)_a]^{-1} & 0 \\
0 & 0 & 0 & [(\partial/\partial x_3)^a(\partial/\partial x_3)_a]^{-1}
\end{array}$$

where $D = (\tau_a\tau^a)(\varphi_b\varphi^b) - (\tau_a\varphi^a)^2$. ($D < 0$ in $M^-$, since $(\varphi_b\varphi^b) < 0$ in $M^-$ and $(\tau_a\tau^a) > 0$ everywhere.)

For future reference, we note that $\nabla^a t$ ($= g^{ab}\nabla_b t = g^{ab}(dt)_b$) can be expressed as

$$\nabla^a t = (\varphi^n\varphi_n)D^{-1}(\partial/\partial t)^a + (-\tau^n\varphi_n)D^{-1}(\partial/\partial\varphi)^a = (D^{-1})\left[(\varphi^n\varphi_n)\tau^a - (\tau^n\varphi_n)\varphi^a\right]$$

in $M^-$. Hence

$$(\nabla_a t)(\nabla^a t) = (\varphi^n\varphi_n)D^{-1}$$

and therefore

$$(\nabla^a t)\left[(\nabla_n t)(\nabla^n t)\right]^{-1} = \tau^a - (\tau^n\varphi_n)(\varphi^m\varphi_m)^{-1}\varphi^a \tag{3}$$

in $M^-$.

Of special interest is the case where $[(\tau^n\varphi_n)(\varphi^m\varphi_m)^{-1}]$ is constant on $M^-$.[14] We will say then that the background spacetime $(M, g_{ab})$ is *static*. This is a slightly non-standard way of formulating the definition.[15] But if $[(\tau^n\varphi_n)(\varphi^m\varphi_m)^{-1}]$ *is* constant, $\tau'^a = \tau^a - (\tau^n\varphi_n)(\varphi^m\varphi_m)^{-1}\varphi^a$ is a smooth timelike Killing field on M that is hyper-surface orthogonal. (It is a Killing field since any linear combination of two Killing fields is one. It is hypersurface orthogonal by (3).)

An example to which we will turn repeatedly is Kerr spacetime (see, e.g., O'Neill (1995)). In Boyer-Lindquist coordinates $(t, \varphi, r, \theta)$ -- again with $\tau^a = (\partial/\partial t)^a$ and $\varphi^a = (\partial/\partial\varphi)^a$ -- the non-zero components are

$$\tau^a\tau_a = 1 - 2Mr\rho^{-2} \tag{5a}$$

$$\tau^a\varphi_a = 2Mra(\sin^2\theta)\rho^{-2} \tag{5b}$$

$$\varphi^a \varphi_a \quad = \quad -[\, r^2 + a^2 + 2\,M\,r\,a^2\,(\sin^2\theta)\,\rho^{-2}\,]\,(\sin^2\theta) \tag{5c}$$

$$(\partial/\partial r)^a (\partial/\partial r)_a \;=\; -\rho^2 \Delta^{-1}$$

$$(\partial/\partial\theta)^a (\partial/\partial\theta)_a \;=\; -\rho^2$$

where

$$\rho^2 = r^2 + a^2 (\cos^2\theta) \tag{5d}$$

$$\Delta = r^2 - 2\,M\,r + a^2. \tag{5e}$$

(Here a and M are positive constants.) Axis points are those at which $(\sin^2\theta) = 0$. It is not the case that $\tau^a$ is everywhere timelike and $\varphi^a$ is everywhere spacelike on $M^-$. But these conditions do obtain in restricted regions of interest, e.g., in the open set where $r > 2\,M$. If we think of Kerr spacetime as representing the spacetime structure surrounding a rotating black hole, our interest will be in small rings that are positioned close to the axis of rotational symmetry (where $(\sin^2\theta)$ is small) and far away from the center point (where r is large). There we can sidestep complexities having to do with horizons and singularities.

## II.2 Striated Orbit Cylinders and the ZAM Criterion

Assume we have fixed, once and for all, a stationary, axi-symmetric spacetime (M, $g_{ab}$) with isometry groups $\{\Gamma_t\colon t\in \mathbf{R}\}$ and $\{\Sigma_\varphi\colon \varphi\in \mathbf{S}^1\}$ and corresponding Killing fields $\tau^a$ and $\varphi^b$. The first of these fields defines a temporal orientation on (M, $g_{ab}$). We will work with that one in what follows.[16]

We want to represent (one-dimensional) rings, centered about the axis of rotational symmetry. We do so using the "orbit cylinders" introduced above. Recall that these were characterized as the orbits of points in $M^-$ under the two–parameter isometry group $\{\Gamma_t \circ \Sigma_\varphi\colon t\in \mathbf{R}\ \&\ \varphi\in \mathbf{S}^1\}$. Here is an equivalent formulation.

<u>Definition</u> An *orbit cylinder* is a smooth, two-dimensional, timelike submanifold in $M^-$, diffeomorphic to the cylinder $\mathbf{R}\times \mathbf{S}^1$, that is invariant under the action of all maps $\Gamma_t$ and $\Sigma_\varphi$.

Clearly, we are thinking about the life history of a ring, not its state at a given "time".

Let C be an orbit cylinder representing ring R. To represent the rotational state of R, we need to keep track of the motion of individual points on it. Each such point has a worldline that can be represented as a timelike curve on C. So we are led to consider, not just C, but C together with a congruence of smooth timelike curves on C.[17]

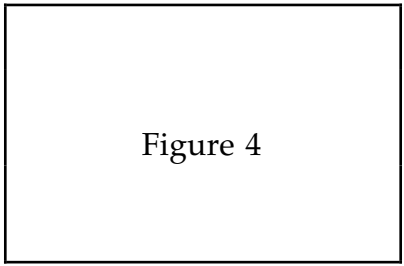

Figure 4

We want to think of the ring as being in a state of rigid rotation (with the distance between points on the ring remaining constant). So we are further led to restrict attention to just those congruences of timelike curves on C that are invariant under all isometries $\Gamma_t$ and $\Sigma_\varphi$. Equivalently (moving from the curves themselves to their tangent vectors), we are led to consider smooth, future directed timelike vector fields $\xi^a$ on C that are invariant under all these maps. Since each such field is determined by its value at any one point on C (and since the tangent plane to C at any point is spanned by the vectors $\tau^a$ and $\varphi^a$ there), $\xi^a$ must be of the form $(k_1 \tau^a + k_2 \varphi^a)$, where $k_1$ and $k_2$ are constants, and $k_1 > 0$.[18] We lose nothing if we rescale $\xi^a$ by a positive factor and write it in the form $(\tau^a + k\,\varphi^a)$. So we are led to the following definition.

<u>Definition</u>  A *striated orbit cylinder* is a pair (C, k), where C is an orbit cylinder,

and k is a number such that the vector field $(\tau^a + k\,\varphi^a)$ is timelike on C.

We call the integral curves of $(\tau^a + k\,\varphi^a)$ on C "striation lines", and call k their "slope factor".

Now we can formulate our fundamental question: *Under what conditions does a striated orbit cylinder count as non-rotating?* The first proposal we consider is the following.[19]

<u>Definition</u>  A striated orbit cylinder (C, k) is non-rotating according to the *ZAM* (*zero angular momentum*) *criterion* if the vector field $(\tau^a + k\varphi^a)$ is orthogonal to $\varphi^a$, i.e., if $k = -(\tau^a \varphi_a)(\varphi^n \varphi_n)^{-1}$.

The connection with angular momentum is immediate. The stated condition is equivalent to the assertion that every point on the ring has 0 angular momentum with respect to the rotational Killing field $\varphi^a$.[20]

The criterion should seem like a reasonable one. It seems plausible to regard a striated orbit cylinder as non-rotating iff the striation lines (representing the worldlines of points on the ring) do not "wrap around the cylinder". And the latter condition is plausibly captured in the requirement that the striation lines be everywhere orthogonal to equatorial circles on the cylinder, i.e., have no component in the direction of those circles.  But how does one characterize "equatorial circles" in the present context? If the background geometry were Euclidean, e.g., if we were dealing with ordinary barber shop poles, we could characterize an equatorial circle as a closed curve of shortest length on the cylinder that is not contractible to a point. That characterization does not carry over to the present context where the background metric has Lorentz signature. But an alternate, equivalent one does. In the Euclidean case, we can equally well characterize an equatorial circle as the orbit of a point under the group of rotations that leave fixed the central axis of the cylinder. Lifting that characterization to the present context, we are led to construe the orbits of points under $\{\Sigma_\varphi : \varphi \in \mathbf{S}^1\}$ as "equatorial circles". These are just the integral curves of the field $\varphi^a$. So the requirement that striation lines not "wrap around the cylinder" is plausibly captured in the condition that they be everywhere orthogonal to the field $\varphi^a$. That is precisely the ZAM criterion of non-rotation.

Consider now the operational test described in the preceding section for whether a ring is rotating according to the ZAM criterion. We can verify that it works with a simple calculation.[21] Let (C, k) be a striated orbit cylinder. We have to keep track of three curves on C. (See figure 5). The first is a striation line γ that represents the worldline of a fixed point on the ring from which light is emitted and absorbed. The other two are null curves $\lambda_1$ and $\lambda_2$ on C that represent the worldlines of photons that start at that point, traverse the ring in opposite directions, and then arrive back at it. (Call them "photon 1" and "photon 2".) Let $p_0$ be the initial emission point at which the three curves intersect. Let $p_1$ be the intersection point of γ with $\lambda_1$ at which the first photon is reabsorbed. And let $p_2$ be the corresponding intersection point of γ with $\lambda_2$. We have to verify that the photons arrive back at the same instant iff (C, k) is non-rotating according to the ZAM criterion, i.e.,

$$p_1 = p_2 \quad \Leftrightarrow \quad k = -(\tau^a \varphi_a)(\varphi^n \varphi_n)^{-1}. \tag{6}$$

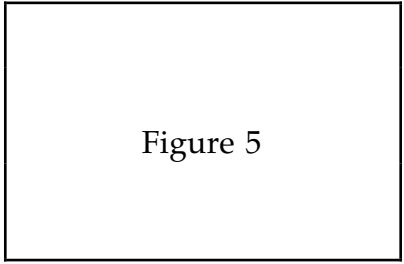

Figure 5

The tangent fields to the curves γ, $\lambda_1$, and $\lambda_2$ (after rescaling by a positive constant) can be written in the form $(\tau^a + k\,\varphi^a)$, $(\tau^a + l_1\,\varphi^a)$, and $(\tau^a + l_2\,\varphi^a)$. Since the first is timelike, and the second two are null, we have $l_i \neq k$ and

$$(\tau^a + l_i\,\varphi^a)(\tau_a + l_i\,\varphi_a) = 0 \tag{7}$$

for $i = 1, 2$. Consider the scalar function $\varphi': C \to \mathbf{S}^1$ defined by $\varphi' = (\varphi - kt) \pmod{2\pi}$. It is a circular coordinate that is adapted to $(C, k)$ in the sense that it is constant along striation lines.[22] Let the $(t, \varphi')$ coordinates of the points $p_0, p_1, p_2$ be $(t_0, \varphi'_0)$, $(t_1, \varphi'_0)$, and $(t_2, \varphi'_0)$. We can verify (6) by considering the respective rates at which $\varphi'$ changes along $\lambda_1$ and $\lambda_2$ as a function of $t$.[23] Without loss of generality, assume that it increases from $\varphi'_0$ to $(\varphi'_0 + 2\pi)$ along $\lambda_1$, and decreases from $\varphi'_0$ to $(\varphi'_0 - 2\pi)$ along $\lambda_2$. Then, the total increase (resp. decrease) along $\lambda_1$ (resp. $\lambda_2$) can be expressed as[24]:

$$2\pi = (t_1 - t_0)(d\varphi'/dt)|_{\text{on } \lambda_1} = (t_1 - t_0)(l_1 - k)$$
$$-2\pi = (t_2 - t_0)(d\varphi'/dt)|_{\text{on } \lambda_2} = (t_2 - t_0)(l_2 - k).$$

So

$$(t_1 - t_2) = 2\pi\,(l_1 + l_2 - 2k)(l_1 - k)^{-1}(l_2 - k)^{-1}.$$

But it follows from (7) that

$$l_1 = [-(\tau^a\varphi_a) + (-D)^{1/2}]\,(\varphi^n\varphi_n)^{-1}$$
$$l_2 = [-(\tau^a\varphi_a) - (-D)^{1/2}]\,(\varphi^n\varphi_n)^{-1}$$

where (as above) $D = (\tau_a\tau^a)(\varphi_b\varphi^b) - (\tau_n\varphi^n)^2$. So,

$$p_1 = p_2 \Leftrightarrow (t_1 - t_2) = 0 \Leftrightarrow (l_1 + l_2 - 2k) = 0 \Leftrightarrow k = -(\tau^a\varphi_a)(\varphi^n\varphi_n)^{-1},$$

which confirms (6).

II.3 Generalized Criteria of Non-Rotation and the Relative Rotation Condition

Now we turn to "generalized criteria of non-rotation". Using our current terminology, the definition comes out this way.

Definition  A *generalized criterion of non-rotation* is a specification, for every striated orbit cylinder $(C, k)$, whether it is to count as "non-rotating" or not.

-18-

We do not assume that generalized criteria of non-rotation bear a natural geometrical or experimental interpretation. Nor do we assume that given an orbit cylinder C, they render (C, k) "non-rotating" for at least one k, or at most one k. Clearly, the ZAM criterion of non-rotation qualifies as a generalized criterion of such.

Next consider the relative rotation condition. Intuitively, it asserts that if we have two rings (with the same axis of symmetry), then if the first qualifies as "non-rotating", and if the second is non-rotating relative to the first, then the second ring also qualifies as "non-rotating". As mentioned above, all we need here is a sufficient condition for relative non-rotation of the two rings; and it seems, at least, a plausible sufficient condition for this that, over time, there be no change in the distance between any point on one ring and any point on the other, i.e., the two rings move as if locked together.

Suppose we have two striated orbit cylinders $(C_1, k_1)$ and $(C_2, k_2)$, suppose $\gamma_1$ is a striation line of the first, and $\gamma_2$ is a striation line of the second. There are various ways we might try to measure the "distance between $\gamma_1$ and $\gamma_2$". For example, we might bounce a photon back and forth between them and keep track of how much time is required for the round trip -- as measured by a clock following one of the striation lines. But no matter what procedure we use, the measured distance will be constant over time if $\gamma_1$ and $\gamma_2$ are (up to reparametrization) integral curves of a common Killing field (or, equivalently, orbits of a common one-parameter group of isometries). For any measurement procedure can be characterized in terms of some set of relations and functions that are definable in terms $g_{ab}$ (e.g., the set of null geodesics, the length of a timelike curve) and *all* such relations and functions will be preserved under the elements of the isometry group (since these all preserve $g_{ab}$). So we seem to have a plausible sufficient condition for the relative non-rotation of ($C_1$, $k_1$) and ($C_2$, $k_2$) -- namely, that there exist a (single) Killing field $\kappa^a$ defined on M whose restriction to $C_1$ is proportional to ($\tau^a + k_1\varphi^a$), and whose restriction to $C_2$ is proportional to ($\tau^a + k_2\varphi^a$). But the latter condition holds immediately if $k_1 = k_2$, since, for any constant k, ($\tau^a + k\varphi^a$) is itself a Killing field defined on M.

The upshot of this long-winded argument is the proposal that it is plausible to regard ($C_2$, $k_2$) as non-rotating relative to ($C_1$, $k_1$) if $k_1 = k_2$. So we are led to the following formulation of the relative rotation condition.

> <u>Relative Rotation Condition</u>  For all k, and all striated orbit cylinders ($C_1$, k) and ($C_2$, k) sharing k as their slope factor, if ($C_1$, k) qualifies as "non-rotating", so does ($C_2$, k).[25]

It follows easily that *the ZAM criterion of non-rotation satisfies the relative rotation condition iff the background stationary, axi-symmetric spacetime structure is static, i.e., if the function $[(\tau^a\varphi_a)(\varphi^n\varphi_n)^{-1}]$ is constant on $M^-$.*[26] In Kerr spacetime, by (5b) and (5c),

$$-(\tau^a\varphi_a)(\varphi^n\varphi_n)^{-1} = (2\,M\,r\,a)\,[\,(r^2 + a^2)\,\rho^2 + 2\,M\,r\,a^2\,(\sin^2\theta)\,]^{-1}. \qquad (8)$$

The right hand side expression is not constant over any open set. So we see that *the ZAM criterion does not satisfy the relative rotation condition in Kerr spacetime, or the restriction of Kerr spacetime to any open set.* (We have been taking for granted that a and M are both strictly positive. It also follows directly from (8) that the ZAM criterion *does* satisfy the relative rotation condition in Schwarzschild spacetime (a = 0 and M > 0) and Minkowksi spacetime (a = 0 and M = 0).)

### II.4  The CIA Criterion

Next we consider how to capture the CIA criterion of non-rotation in the language of spacetime geometry. Let (C, k) be a striated orbit cylinder. The Killing field $(\tau^a + k\varphi^a)$ that determines the striation lines on C is defined on all of M. It seems a natural proposal to construe (C, k) as non-rotating if the twist (or vorticity) of $(\tau^a + k\varphi^a)$ vanishes at axis points. This is very close to being the CIA criterion. But there is a problem. It is true that given any axis point p, there is one and only k such that $(\tau^a + k\varphi^a)$ has vanishing twist at p. (We verify this in lemma 2.) But it turns out that that critical value need not be the same at all axis points. It is not, for example, in Kerr spacetime. (In the end, it is this one fact that lies at the heart of our mini no-go theorem.) So we need to direct attention to some particular axis point and take the test to be whether $(\tau^a + k\varphi^a)$ has vanishing twist *there*. The natural choice is the "centerpoint" of the ring, the point that lies at "the intersection of the axis with the plane of C". (That is where we previously placed the experimental apparatus consisting of lazy susan + water-bucket + telescope. Recall figure 2.) The question, then, is how to construe the expressions in quotation marks.

One natural way to do so is in terms of light signals traveling from the ring to the axis. In cases of interest, there is exactly one point on the axis at which the incoming light signals arrive so as to be perpendicular to the axis. That one point is a natural candidate for the "centerpoint" of the ring, and we will treat it as such in what follows. But a bit of work is necessary to set everything up.

Let $\epsilon^{abcd}$ be a volume element[27], and let $\sigma^a$ be the smooth field defined by
$$\sigma^a = \epsilon^{abcd} \tau_b \nabla_c \varphi_d.$$
We claim that at every axis point p, $\sigma^a$ gives the "direction of the axis of rotation" as determined relative to $\tau^a$. The interpretation is supported by the following lemma that collects several simple facts about $\sigma^a$ for future reference. It implies that at axis points, $\sigma^a$ is, up to a constant, the only non-zero vector, orthogonal to $\tau^a$, that is kept invariant by all isometries $\Sigma_\varphi$.

Lemma 1  At all points:

(i) $\sigma^a$ is orthogonal to $\tau^a$ and $\varphi^a$

(ii) $\mathbf{L}_\varphi(\sigma^a) = \mathbf{0} = \mathbf{L}_\tau(\sigma^a)$     (Here $\mathbf{L}_\varphi$ and $\mathbf{L}_\tau$ are Lie derivative operators.)

(iii) $\tau_{[a} \nabla_b \varphi_{c]} = (1/6) \, \epsilon_{abcd} \sigma^d$.

At axis points:

(iv) $\sigma^a \neq \mathbf{0}$

(v) $\nabla_a \varphi_b = (1/2)(\tau^n \tau_n)^{-1} \epsilon_{abcd} \tau^c \sigma^d$

Given any field $\psi^a$, if $\mathbf{L}_\varphi(\psi^a) = \mathbf{0}$ at an axis point, then at the point it must be of form $\psi^a = k_1 \tau^a + k_2 \sigma^a$.

Proof  (i) $\epsilon^{abcd}$ is totally anti-symmetric. So $\sigma^a \tau_a = \epsilon^{abcd} \tau_a \tau_b \nabla_c \varphi_d = 0$, and

$$\sigma^a \varphi_a = \epsilon^{abcd} \varphi_a \tau_b \nabla_c \varphi_d = \epsilon^{abcd} \varphi_{[a} \tau_b \nabla_c \varphi_{d]}.$$

But $\varphi_{[a} \tau_b \nabla_c \varphi_{d]} = \mathbf{0}$, by (2b). So $\sigma^a \varphi_a = 0$.  (ii) The Lie operators $\mathbf{L}_\tau$ and $\mathbf{L}_\varphi$ annihilate $\tau^a$ and $\varphi^a$, by (1), and annihilate $g_{ab}$ and $\epsilon^{abcd}$ because $\tau^a$ and $\varphi^a$ are Killing vectors. So they annihilate all fields definable in terms of $\tau^a$, $\varphi^a$, $g_{ab}$, and $\epsilon^{abcd}$, including $\sigma^a$. (iii) follows by a simple computation:

$$\epsilon_{abcd} \sigma^d = \epsilon_{abcd} \epsilon^{dmpq} \tau_m \nabla_p \varphi_q = (3!) \, \delta_a^{[m} \delta_b^{p} \delta_c^{q]} \tau_m \nabla_p \varphi_q = 6 \, \tau_{[a} \nabla_b \varphi_{c]}.[28]$$

For (iv), suppose $\sigma^a = \mathbf{0}$ at an axis point p. Then, by (iii), $\tau^a \tau_{[a} \nabla_b \varphi_{c]} = \mathbf{0}$ at p. Expanding this equation and using the fact that $\nabla_b \varphi_c = -\nabla_c \varphi_b$ (since $\tau^a$ is a Killing field), we have

$$(\tau^a \tau_a) \nabla_b \varphi_c + \tau_c \tau^a \nabla_a \varphi_b - \tau_b \tau^a \nabla_a \varphi_c = \mathbf{0}.$$

But the second and third terms are $\mathbf{0}$ at p, since $\tau^a \nabla_a \varphi_b = \varphi^a \nabla_a \tau_b$ (by equation (1)) and $\varphi_b = \mathbf{0}$ at p. So, since $\tau^a$ is timelike, $\nabla_b \varphi_c = \mathbf{0}$ at p. But this is impossible. For given any Killing field $\kappa^a$, if both $\kappa_a$ and $\nabla_a \kappa_b$ vanish at a point, $\kappa^a$ must vanish everywhere. (See Wald (1984), page 443.) And we know that $\varphi^a$ does not vanish everywhere. (v) follows from (iii) and a computation very close to the one just used for (iv). Finally, assume that $\mathbf{L}_\varphi(\psi^a) = \mathbf{0}$ at p. Then, at p, $\psi^a \nabla_a \varphi_b = \varphi^a \nabla_a \psi_b = \mathbf{0}$ since $\varphi^a = 0$ at p. Hence, by (v), $\epsilon_{abcd}\psi^a\tau^c\sigma^d = 0$ at p. It follows that the three vectors $\psi^a$, $\tau^a$, and $\sigma^a$ are linearly dependent at p and, so, $\psi^a$ can be expressed there as a linear combination of the other two vectors. ♦

Let C be an orbit cylinder. Let γ be an integral curve of $\tau^a$ on which $\varphi^a$ vanishes.[29] It represents the worldline of a point[30] on the axis of rotation. We say that γ is the *centerpoint* of C if, for all future-directed null geodesics running from a point on C to a point on γ, if $\lambda^a$ is the tangent field to the null geodesic, then, at the latter (arrival) point, $\lambda^a$ is orthogonal to $\sigma^a$.[31]

In what follows, we take for granted that orbit cylinders *have* unique centerpoints. The assumption is harmless because it will suffice for our purposes to restrict attention to regions of spacetime near axis points (e.g., within convex sets) and there they certainly do.[32]

To complete our definition of the CIA criterion we need the following lemma.

<u>Lemma 2</u> Let p be an axis point. Then there is a unique k such that the Killing field $\xi^a = (\tau^a + k\varphi^a)$ has vanishing twist at p, i.e., such that $\xi_{[a} \nabla_b \xi_{c]} = 0$ at p. It's value is given by:
$$k_{crit}(p) = -[(\nabla_b \tau_c)(\nabla^b \varphi^c)] [(\nabla_m \varphi_n)(\nabla^m \varphi^n)]^{-1}.$$

<u>Proof</u> Since $\varphi^a = 0$ at p, what we need to show is that there is a unique k such that
$$\tau_{[a} \nabla_b \tau_{c]} + k \tau_{[a} \nabla_b \varphi_{c]} = 0 \qquad (9)$$
at p. We know from clauses (iii) and (iv) of lemma 1 that $\tau_{[a} \nabla_b \varphi_{c]} \neq 0$ at p. So uniqueness is immediate. For existence, consider the twist vector field of $\tau^a$ defined by
$$\omega^a = \epsilon^{abcd} \tau_b \nabla_c \tau_d.$$

$\omega^a$ is orthogonal to $\tau^a$ and is Lie derived by $\varphi^a$, i.e., $\mathbf{L}_\varphi(\omega^a) = \mathbf{0}$. (The proof is almost exactly the same as for $\sigma^a$ in clauses (i) and (ii) of lemma 1.) Hence, by the final assertion in lemma 1, $\omega^a = k_2 \sigma^a$ at p, for some number $k_2$. It follows by clause (iii) of lemma 1, and the counterpart statement for $\omega^a$ and $\tau_{[a}\nabla_b \varphi_{c]}$, that

$$\tau_{[a}\nabla_b \tau_{c]} = (1/6)\,\epsilon_{abcd}\omega^d = k_2(1/6)\,\epsilon_{abcd}\sigma^d = k_2\,\tau_{[a}\nabla_b \varphi_{c]}.$$

Thus (9) will be satisfied if we take $k = -k_2$.

Now assume that k *does* satisfy (9) at p. Contracting with $\tau^a\nabla^b\varphi^c$, and then dividing by $(\tau^c \tau_c)$, yields

$$[(\nabla_b \tau_c)(\nabla^b\varphi^c)] + k\,[(\nabla_b\varphi_c)(\nabla^b\varphi^c)] = 0.$$

(Almost all terms drop out because $\tau^a\nabla_a\varphi_b = 0$.) So to complete the proof we need only verify that $(\nabla_b \varphi_c)(\nabla^b\varphi^c) \neq \mathbf{0}$ at p. But this follows, since by clause (v) of lemma 1,

$$(\nabla_b\varphi_c)(\nabla^b\varphi^c) = (1/2)(\tau^n\tau_n)^{-1}\epsilon_{bcpq}(\nabla^b\varphi^c)\tau^p\sigma^q = -(1/2)(\tau_n\tau^n)^{-1}(\sigma_q\sigma^q)$$

at p, and by clauses (i) and (iv), $(\sigma_q\sigma^q) < 0$ at p. ♦

Lemma 2 has a simple geometric interpretation. Equation (9) is equivalent to:

$$\omega^a + k\sigma^a = \mathbf{0}.$$

So, when the dust clears, the lemma asserts that, at every axis point, the twist vector $\omega^a$ (of $\tau^a$) is co-alligned with the axis direction vector $\sigma^a$. The critical value k is just a proportionality factor.

If p is an axis point, and $\gamma$ is the integral curve of $\tau^a$ that passes through p, the function $k_{crit}$ is constant on $\gamma$. (This follows since the condition that $(\tau^a + k\varphi^a)$ is twist free is definable in terms of $\tau^a$, $\varphi^a$, and $g_{ab}$, and these are all preserved by the isometries $\Gamma_t$). So, in particular, if $\gamma$ is the centerpoint of an orbit cylinder C, we can write '$k_{crit}(\gamma)$' without ambiguity. Now we have all the pieces in place for our definition.

<u>Definition</u>  A striated orbit cylinder (C, k) is non-rotating according to the *CIA*

*criterion* if k = $k_{crit}(\gamma)$, where $\gamma$ is the centerpoint of C.

We now consider the conditions under which the CIA criterion satisfies the relative rotation condition, and the conditions under which our two criteria of non-rotation agree. We take them in order.[33] Since every axis point is the centerpoint of some orbit cylinder, the CIA criterion satisfies the relative rotation condition iff the function $k_{crit}$ assumes the same value at all axis points. But there is a more instructive way to formulate the later condition.

<u>Lemma 3</u>  The function $f = [-(\tau^a \varphi_a)(\varphi^n \varphi_n)^{-1}]$ can be smoothly extended from $M^-$ to all of M.  The value of the extension at an axis point p is $k_{crit}(p)$.

<u>Proof</u>  The proof that f can be smoothly extended to M is long, and we omit the details.[34]  But the rest of the proof is easy.  Consider the field $\tau'^a = \tau^a + f\varphi^a$ defined on $M^-$. By (3), it is hypersurface orthogonal, i.e., of form $\tau'_a = g \nabla_a h$. So, it must have vanishing twist.[35] Therefore, at all points in $M^-$,

$$0 = \tau'_{[a} \nabla_b \tau'_{c]} = \tau_{[a} \nabla_b \tau_{c]} + f \tau_{[a} \nabla_b \varphi_{c]} + \varphi_{[c} \tau_a \nabla_{b]} f + f \varphi_{[a} \nabla_b \tau_{c]} + f^2 \varphi_{[a} \nabla_b \varphi_{c]}.$$

Let k' be the limiting value of f at p.  Then, at p we have

$$0 = \tau_{[a} \nabla_b \tau_{c]} + k' \tau_{[a} \nabla_b \varphi_{c]}$$

(since $\varphi_a = 0$  at p).  But we saw in the proof of lemma 3 that there is a unique k that satisfies equation (9). So $k' = k_{crit}(p)$. ♦

It follows immediately from lemma 3 that *the CIA criterion satisfies the relative rotation condition iff $[-(\tau^a \varphi_a)(\varphi^n \varphi_n)^{-1}]$ has the same limit values at all axis point*. At axis points in Kerr spacetime, the limit value of $[-(\tau^a \varphi_a)(\varphi^n \varphi_n)^{-1}]$ is

$$2 M r a [r^2 + a^2]^{-2}.$$

(Recall (8).) Clearly, this function is not constant over any interval of values for r. So we see that *the CIA criterion does not satisfy the relative rotation condition in Kerr spacetime, or the restriction of Kerr spacetime to an open set containing an axis*

*point.*

We also see if that *if the background stationary, axi-symmetric spacetime is static, then the CIA criterion satisfies the relative rotation condition.* (If $[-(\tau^a\varphi_a)(\varphi^n\varphi_n)^{-1}]$ is constant, then certainly the function has the same limit values at all axis point.) It turns out, however, that *the converse is false.*[36]

Next consider the conditions under which our two criteria agree (for all rings). What is required is that, for all orbit cylinders C, the value of $[-(\tau^a\varphi_a)(\varphi^n\varphi_n)^{-1}]$ on C be equal to the value of $k_{crit}$ at the centerpoint of C. Recalling how centerpoints are defined, and making use of lemma 3, we see that *the CIA and ZAM criteria of non-rotation agree (for all rings) iff the function $[-(\tau^a\varphi_a)(\varphi^n\varphi_n)^{-1}]$ is constant on all null geodesics that terminate at axis points and have tangents there orthogonal to the axis direction $\sigma^a$.*[37] It follows that *they do not agree in Kerr spacetime, or any open set in Kerr spacetime containing an axis point.*[38]

## II.4 The Limit Condition and the Theorem

Finally, we turn to the limit condition. Let $(C_1, k_1)$, $(C_2, k_2)$, $(C_3, k_3)$, ... be a sequence of striated orbit cylinders that share a common centerpoint $\gamma$, and that converge to $\gamma$. (We can take the second condition to mean that each point on $\gamma$ is the accumulation point of a sequence of points $p_1, p_2, p_3, ...$ , with $p_i$ on $C_i$.) For all i, let $\tau_i^a = \tau^a + k_i \varphi^a$, and let $\omega_i^a$ be its associated twist field

$$\omega_i^a = \epsilon^a{}_{bcd} \tau_i^b \nabla^c \tau_i^d.$$

We can take the limit condition to assert that, if each $(C_i, k_i)$ qualifies as "non-rotating", then the sequence $\omega_1^a, \omega_2^a, \omega_3^a, ...$ converges to 0 on $\gamma$. This captures the requirement that the measured angular velocity of $(C_i, k_i)$ relative to the compass of inertia on $\gamma$ goes to 0. An equivalent formulation is the following.[39]

<u>Limit Condition</u>  Let $(C_1, k_1)$, $(C_2, k_2)$, $(C_3, k_3)$, ... be a sequence of striated orbit cylinders that share a common centerpoint $\gamma$, and that converge to $\gamma$. If each of the $(C_i, k_i)$ qualifies as "non-rotating", then $\lim_{i \to \infty} k_i = k_{crit}(\gamma)$.

It follows immediately, of course, that *the CIA criterion satisfies the limit condition* (in all stationary, axi-symmetric spacetime models). For if each $(C_i, k_i)$ qualifies as non-rotating according to that criterion, $k_i = k_{crit}(\gamma)$ for all i. (One does not need to take a limit to reach $k_{crit}(\gamma)$.) It also follows immediately from lemma 3 that *the ZAM criterion satisfies the limit condition* (in all stationary, axi-symmetric spacetime models). For if each $(C_i, k_i)$ qualifies as non-rotating according to that criterion, $k_i$ is equal to the value of the function $[-(\tau^a \varphi_a)(\varphi^n \varphi_n)^{-1}]$ on $C_i$, for all i. And the sequence of *those* values converges to $k_{crit}(\gamma)$ by the lemma (and the fact that the $C_i$ converge to $\gamma$).

We can, now, finally, state our principal result.

<u>Theorem</u>  Assume the background stationary, axi-symmetric spacetime model is one (like Kerr spacetime) in which there exist axis points p and p' such that $k_{crit}(p) \neq k_{crit}(p')$. Then there is no generalized criterion of non–rotation that (in the model) satisfies the following three conditions:
  (i)   limit condition
  (ii)  relative rotation condition
  (iii) non-vacuity condition:  there is at least one striated orbit cylinder
        that qualifies as "non-rotating".

<u>Proof:</u> Let $\gamma$ and $\gamma'$ be the integral curves of $\tau^a$ containing p and p', and let $C_1, C_2, C_3,$ ... and $C'_1, C'_2, C'_3,$ ... be sequences of orbit cylinders that converge to $\gamma$ and $\gamma'$ respectively. Now assume there is a generalized criterion of non-rotation G that satisfies all three conditions in the model. Let (C, k) be a striated orbit cylinder that

-27-

qualifies as "non-rotating" according to G. For all i sufficiently large, $(C_i, k)$ and $(C'_i, k)$ are striated orbit cylinders, i.e., $(\tau^a + k\varphi^a)$ is timelike on $C_i$ and $C'_i$. So, by the relative rotation condition, $(C_i, k)$ and $(C'_i, k)$ qualify as non-rotating according to G for all i sufficiently large. Therefore, by the limit condition applied to the sequences $(C_1, k), (C_2, k), (C_3, k), ...$ and $(C'_1, k), (C'_2, k), (C'_3, k), ...$, it must be the case that $k = k_{crit}(p)$ and $k = k_{crit}(p')$. But this contradicts our hypothesis that $k_{crit}(p) \neq k_{crit}(p')$. So our non-existence claim follows. ♦

The implication in the theorem is reversible. For if the value of $k_{crit}$ *is* the same at all axis points, then the CIA criterion satisfies all three of the stated conditions. (Even then, of course, it need not be the case that the CIA criterion agrees with ZAM criterion, or that the latter satisfies the relative rotation condition.)

...................................................

I have argued that, in the context of general relativity, the concept of rotation is a delicate and interesting one. Perhaps it is worth saying, in conclusion, that I intend no stronger claim. There is no suggestion here that the no-go theorem poses a deep interpretive problem (or any problem at all) for the foundations of general relativity, nor that we have to give up talk about rotation in general relativity. The point is just that, depending on the circumstances, we may have to disambiguate different criteria of rotation, and may have to remember that they all leave our classical intuitions far behind.

# Footnotes

[1] It is a pleasure to dedicate this paper to Howard Stein. It has been one of the great good fortunes of my life to be able to count him as my teacher, colleague, and friend for over twenty-five years. I wish to thank him for the assistance and support he has given me during this time, and to thank him, David Garfinkle, Robert Wald, and, especially, Robert Geroch for helpful discussion of the issues raised in the paper. I am also grateful to John Norton for help in preparing figures 4 and 5.

[2] For an extended discussion of other criteria, see Page (1998) and the references cited there.

[3] It turns out that a generalized criterion of rotation can satisfy both the limit and relative rotation conditions vacuously if, according to the criterion, no ring, in *any* state of motion (or non-motion), qualifies as "non-rotating".

[4] Here and throughout section I, we make free appeal to our common sense (Euclidean) intuitions about the geometry of space. We take for granted that we understand, for example, what it means to say that the plane of the ring is orthogonal to the axis, that the axis is at the center of the ring, etc.. Later, in section II, we will have to consider how to capture these conditions within the framework of four-dimensional spacetime geometry.

[5] We will later restrict attention to spacetimes that are stationary and axi-symmetric. It is the presence of the latter axial (or rotational) symmetry that gives rise to a notion of angular momentum. (See footnote 20.)

[6] I am only thinking here of experimental procedures that can be performed locally, on or near the ring and axis. Procedures performed, for example, at "spatial

infinity", are excluded.

[7] For example, we can take an arbitrary ring in an arbitrary state of uniform rotational motion and dub it "non-rotating". Then we can take other rings to be "non-rotating" precisely if they are non-rotating relative to *that* one (in the sense described).

[8] In what follows, we presuppose familiarity with the basic mathematical formalism of general relativity, and make use of the so-called "abstract index notation" (see Wald (1984)).

[9] Thus, $\Gamma_0$ and $\Sigma_0$ are the identity map on M, and

$$\Gamma_t \circ \Gamma_{t'} = \Gamma_{(t + t')} \quad \text{and} \quad \Sigma_\varphi \circ \Sigma_{\varphi'} = \Sigma_{(\varphi + \varphi') \, (\text{mod } 2\pi)}$$

for all t, t' in **R**, and all $\varphi, \varphi'$ in $\mathbf{S}^1$.

[10] Strictly speaking, this condition rules out standard examples of interest, including Kerr spacetime. We are, in effect, limiting attention to restricted regions of interest in those spacetimes where the condition holds. (See the final paragraph of this section.)

[11] That is, at every point there is a timelike vector tangent to the submanifold. Equivalently, the restriction of $g_{ab}$ to the submanifold has signature (1, –1).

[12] For a proof of the equivalence, see Wald (1984), p. 163.

[13] See the discussion in Wald (1984), pp. 162 -165.

[14] Note that the definition does not depend on the initial choice of timelike Killing field $\tau^a$ in this sense: given any other choice $\tau^{*a} = (k_1 \tau^a + k_2 \varphi^a)$, $\tau^{*a}$ satisfies the constancy condition iff $\tau^a$ does.

[15] Usually one says that a spacetime is "static" if there exists a timelike Killing field $\kappa^a$ (defined everywhere or, at least in some domain of interest) that is hypersurface orthogonal, i.e., such that $\kappa_a = f(\nabla_a g)$ for some functions f and g. (In this case $\kappa^a$ is orthogonal to the g = constant hypersurfaces.)

[16] That is, a timelike vector $\alpha^a$ at a point will qualify as future directed if $\alpha^a \tau_a > 0$.

[17] In what follows, we will not always bother to distinguish between (parametrized) curves and the images of such curves. Strictly speaking, it is usually the latter in which we are interested.

[18] Since $\xi^a$ is timelike and future directed, it must be the case that

$$(k_1 \tau^a + k_2 \varphi^a)(k_1 \tau_a + k_2 \varphi_a) > 0 \quad \text{and} \quad \tau^a (k_1 \tau_a + k_2 \varphi_a) > 0.$$

These two conditions imply that $k_1 > 0$.

[19] Rings non-rotating according to this criterion might also be called "locally non-rotating". That terminology is often used in the literature. (See, for example, Bardeen (1970), p. 79, and Wald (1984), p. 187.)

[20] Given any Killing field $\kappa^a$ in any relativistic spacetime model (not necessarily stationary and axi-symmetric), and any timelike curve with (normalized) four-velocity $\xi^a$, we associate with the two a scalar field $(\kappa^a \xi_a)$ on the curve. If the curve represents a point particle, then we call $(\kappa^a \xi_a)$ the "energy" of the particle (relative to $\kappa^a$) if $\kappa^a$ is timelike, and call it the "angular momentum" of the particle (relative to $\kappa^a$) if $\kappa^a$ corresponds to a rotational symmetry. In the special case of a free particle with geodesic worldline, the canonically associated magnitude $(\kappa^a \xi_a)$ is constant on the curve (i.e., is conserved) since

$$\xi^n \nabla_n (\kappa^a \xi_a) = \kappa^a \xi^n \nabla_n \xi_a + \xi^a \xi^n \nabla_n \kappa_a = \mathbf{0}.$$

(The first term in the sum vanishes because the curve is a geodesic ($\xi^n \nabla_n \xi^a = 0$); the second does so because $\kappa^a$ is a Killing field $\nabla_{(n} \kappa_{a)} = 0$.)

In the case at hand, we are considering a rotational Killing field $\varphi^a$ and points on the ring with four-velocity $f(\tau^a + k\varphi^a)$, where $f = [(\tau^a + k\varphi^a)(\tau_a + k\varphi_a)]^{-1/2}$. The angular momentum of the points (with respect to $\varphi^a$) is $f(\tau^a + k\varphi^a)\varphi_a$. Clearly, this magnitude vanishes precisely if $(\tau^a + k\varphi^a)$ is orthogonal to $\varphi_a$.

[21] See also Bardeen (1970) and Ashtekar and Magnon (1975). Ours is a simple, low-brow calculation. The discussion in the second reference is much more general and insightful. (Readers may want to skip the calculation. It is not needed for anything that follows.)

[22] To confirm that it is constant along them, note that

$$(\tau^a + k\varphi^a)\nabla_a(\varphi - kt) = \tau^a \nabla_a(-kt) + (k\varphi^a)\nabla_a \varphi = 0.$$

[23] t increases along all three curves since

$$(\tau^n + k\varphi^n)\nabla_n t = (\tau^n + l_i \varphi^n)\nabla_n t = 1$$

for $i = 1, 2$.

[24] Note that if $\lambda_i$ is parametrized by s, then

$$(d\varphi'/dt) = (d\varphi'/ds)/(dt/ds) = [(\tau^n + l_i \varphi^n)\nabla_n \varphi'] [(\tau^n + k\varphi^n)\nabla_n t]^{-1} = (l_i - k).$$

[25] It might seem preferable to state the condition this way. For all striated orbit cylinders (C, k) and all orbit cylinders C', if (C, k) qualifies as "non-rotating", then so does (C', k). But there is a problem with this formulation. It takes for granted that (C', k) is a striated orbit cylinder in the first place, i.e., that the field $(\tau^a + k\varphi^a)$ is timelike on C'.

[26] (Recall our slightly non-standard definition of "static" in section II.1.) The "if"

half of the proof is straightforward. The proof of converse involves one small complication. Here is the argument in detail. Assume that the ZAM criterion satisfies the relative rotation condition. Let $p_1$ and $p_2$ be any points in $M^-$, let $C_1$ and $C_2$ be the orbit cylinders that contain them, and let $k_1$ and $k_2$ be the values of the function $[-(\tau^a \varphi_a)(\varphi^a \varphi_a)^{-1}]$ at $p_1$ and $p_2$. We must show that $k_1 = k_2$. We don't know (initially) that either $(C_1, k_2)$ or $(C_2, k_1)$ qualifies as a striated orbit cylinder. But, by moving sufficiently close to the axis, we can find a point $p_3$ such that, if $C_3$ is the orbit cylinder that contains $p_3$, $(C_3, k_1)$ and $(C_3, k_2)$ both qualify as striated orbit cylinders. (For *any* value of k, the vector field $(\tau^a + k\varphi^a)$ is timelike at points sufficiently close to axis points.) $(C_1, k_1)$ and $(C_2, k_2)$ are both non-rotating according to the ZAM criterion. So, by the relative rotation condition, $(C_3, k_1)$ and $(C_3, k_2)$ are both non-rotating according to that criterion. So $k_1$ and $k_2$ must both be equal to the value of $[-(\tau^a \varphi_a)(\varphi^a \varphi_a)^{-1}]$ at $p_3$. Therefore, $k_1 = k_2$.

[27] Volume elements always exist locally, and that is sufficient for our purposes.

[28] For facts such as $\epsilon_{abcd} \epsilon^{dmpq} = (3!) \delta_a^{[m} \delta_b^p \delta_c^{q]}$, see Wald (1984), p. 433.

[29] If $\varphi^a$ vanishes at one point on an integral curve of $\tau^a$, it necessarily vanishes at all points. This follows from equation (1).

[30] There is a certain ambiguity in terminology here. We have taken an "axis point" to be a point in M at which $\varphi^a = \mathbf{0}$. But here we have in mind an "axis point" in the sense of figure 2 (i.e., a point in a three-dimensional space). It is represented by a timelike curve in M. In what follows, when referring to "axis points", it should be clear from context (and notation) which is intended.

[31] Note that if the stated condition holds for one future-directed null geodesic

running from a point on C to a point on γ, it holds for all. For the entire class of such null geodesics is generated from any one under the action of the isometry group $\{\Gamma_t \circ \Sigma_\varphi : t \in \mathbf{R}\ \&\ \varphi \in \mathbf{S}^1\}$. Note too that the requirement that $\lambda^a$ be orthogonal to $\sigma^a$ at the arrival point is equivalent to the (slightly more intuitive) requirement that, at that point, the component of $\lambda^a$ orthogonal to $\tau^a$ (representing the "spatial direction" of the incoming light signal relative to $\tau^a$) be orthogonal to $\sigma^a$.

[32] Here is a rough sketch of the proof. Suppose p is an axis point and suppose $\lambda^a$ is a past–directed null vector at p that is orthogonal to the axis direction $\sigma^a$. We can extend $\lambda^a$ to a past directed null geodesic. Let q be any point on that geodesic and let C(q) be the orbit cylinder that contains q, i.e., the orbit of q under the isometry group $\{\Gamma_t \circ \Sigma_\varphi : t \in \mathbf{R}\ \&\ \varphi \in \mathbf{S}^1\}$. Then, "by construction", C has a centerpoint γ (with p on it).

There is a smooth two-dimensional timelike submanifold S through p that consists entirely of axis points. (At every point of S, the tangent plane to S is spanned by $\tau^a$ and $\sigma^a$, where $\sigma^a$ is as in the preceding paragraph.) If we let $\lambda^a$ range over all past–directed null vectors at points of S that are orthogonal to $\sigma^a$, and consider all points q on the past-directed null geodesics they determine (or at least all such points sufficiently close to p), we sweep out an open set O. The argument in the preceding paragraph shows that every orbit cylinder through every point in O has a centerpoint. Uniqueness follows from that fact that, at least locally, given any point q, there is a unique point p on S such that there is a future directed null geodesic that runs from q to p and whose tangent vector at q is orthogonal to $\sigma^a$.

[33] To support the operational interpretation of the CIA criterion presented in section I, one can proceed as follows. Let (C, k) be a striated orbit cylinder, let γ be the centerpoint of C, and let γ' be a striation line on the cylinder, representing the point on the ring (say R) at which a light source is mounted. (So, both γ and γ' are integral

curves of the field $\tau'^a = \tau^a + k\varphi^a$.) Finally, let $\lambda^a$ be a future directed null geodesic field, the integral curves of which run from $\gamma'$ to $\gamma$. (The latter represent light signals emitted at R and received at the center point.) The entire field of integral curves is generated from any one of them under the action of the isometry group associated with $\tau'^a$, i.e., the field $\lambda^a$ is Lie derived by $\tau'^a$. Suppose the telescope at the center point is tracking the light source. Then the direction of the telescope (as determined by the observer with worldline $\gamma$) is represented by a vector field $v^a$ on $\gamma$ whose value at any point is the component of $\lambda^a$ orthogonal to $\tau'^a$ at that point. It is not difficult to check that the Fermi derivative of $v^a$ along $\gamma$ vanishes iff $k = k_{crit}(\gamma)$. (For details, see the discussion in Malament (forthcoming). The vanishing of that Fermi derivative serves as a surrogate here for the flatness of the water surface in the bucket.)

[34] Here is a rough sketch of a proof (shown to us by Robert Geroch). Let S be the two-dimensional submanifold of axis points. Let $\alpha = (-\tau^a \varphi_a)$ and $\beta = (\varphi^a \varphi_a)$, so that f = $\alpha/\beta$. Let p be a point on S. Given any point q sufficiently close to p, it has a unique orthogonal "projection" q' on S, i.e., there is a unique point q' on S with the property that the geodesic segment running from q' to q is orthogonal to S. So the point q is uniquely distinguished by a pair of objects: (i) the value of $\beta$ at q, and (ii) its orthogonal projection q' on S. Thus, if we restrict attention to a suitable open neighborhod of p, we can think of $\alpha$ as a function defined on a subset A of the product manifold (with boundary) $[0,\infty) \times \mathbf{R}^2$. We first show that $\alpha$ is smooth, not just as a function on M, but also when construed this way (as a function on A). To do so, we consider a finite Taylor series expansion of $\alpha$, up to order n, at p, with partial derivatives taken in directions tangent to, and orthogonal to, S. Since $\alpha$ is constant on orbits of $\varphi^a$, the coefficients in the series, i.e, the mixed partial derivatives of $\alpha$ at p, have a special, simple structure. Those of odd order in

directions orthogonal to S must be 0, and those of even order in those directions can be expressed in terms of derivatives in any one orthogonal direction (and directions tangent to S). This allows us to reinterpret the series as a finite Taylor series expansion (at p) of α *construed as a function on A*.

Next we observe that $\alpha = 0$ and $\nabla_a \alpha = \mathbf{0}$ at p. (The second equation can be proved using clause (v) of lemma 1.) It follows that the terms in the expansion of 0th order in β are 0. So we can divide by β and generate a finite Taylor series expansion for $f = \alpha/\beta$ at p. Since the number of terms n in the original expansion was arbitrary, so is the number of terms in the derived expansion. It follows that all partial derivatives of f (as a function on M) exist and are continuous at p.

[35] The computation is straight forward.

$$\tau'_a \nabla_b \tau'_c = g(\nabla_a h)\nabla_b(g(\nabla_c h)) = g^2(\nabla_a h)(\nabla_b \nabla_c h) + g(\nabla_a h)(\nabla_b g)(\nabla_c h).$$

So, since, $(\nabla_{[b} \nabla_{c]} h) = \mathbf{0}$, and $(\nabla_{[a} h)(\nabla_{c]} h) = \mathbf{0}$, it follows that $\tau'_{[a} \nabla_b \tau'_{c]} = \mathbf{0}$.

[36] For a counterexample, it suffices to find a stationary axi-symmetric spacetime that is not static, but exhibits "cylindrical symmetry", i.e., in which the axis direction field $\sigma^a$ is a Killing field. For the latter condition will guarantee that the function $k_{crit}$ is constant as one moves along the axis. Gödel spacetime is one such. (In terms of standard t, φ, r, y coordinates, $\sigma^a$ turns out to be (up to a constant) just the translational Killing field $(\partial/\partial y)^a$.) (For a description of Gödel spacetime, see Hawking and Ellis (1973). For further discussion of rotation in the model, see Malament (forthcoming).)

[37] It would be nice to have a simpler or more instructive characterization. (I do not have one.)

[38] One can verify this with a calculation, but there is a painless way to see that the

stated constancy condition cannot hold. Start at a point p on the axis with positive r coordinate, choose a future directed null vector $\lambda^a$ at p orthogonal to $\sigma^a$, and consider the (maximally extended) null geodesic through p that has tangent $\lambda^a$ at p. It comes in from "past infinity" where, asymptotically, the value of $[-(\tau^a \varphi_a)(\varphi^n \varphi_n)^{-1}]$ is 0. (Recall (8).) Since its value at p is *not* 0, the function cannot be constant on the geodesic.

[39] By lemma 2, $\tau_{[b} \nabla_c \tau_{d]} + k_{crit} \tau_{[b} \nabla_c \varphi_{d]} = \mathbf{0}$ on $\gamma$. So

$\{\omega_i{}^a\}$ converges to 0 on $\gamma$

$\Leftrightarrow \{\tau_{[b} \nabla_c \tau_{d]} + k_i \tau_{[b} \nabla_c \varphi_{d]}\}$ converges to $\mathbf{0}$ on $\gamma$

$\Leftrightarrow \{(k_{crit} - k_i) \tau_{[b} \nabla_c \varphi_{d]}\}$ converges to $\mathbf{0}$ on $\gamma$.

Since $\tau_{[b} \nabla_c \varphi_{d]} \neq \mathbf{0}$ on $\gamma$, the third conditions holds iff $(k_{crit} - k_i)$ converges to 0.

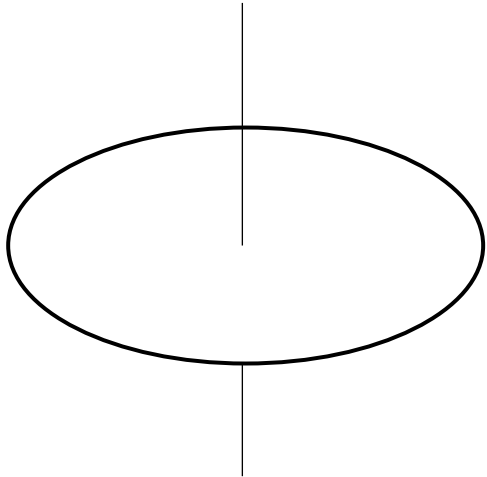

Figure 1

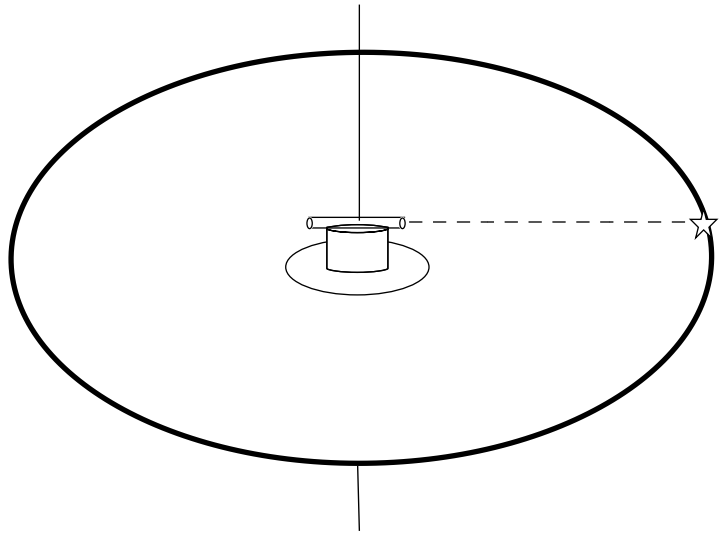

Figure 2

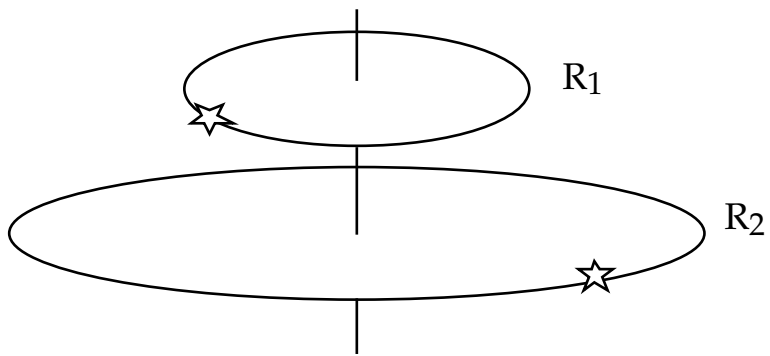

Figure 3

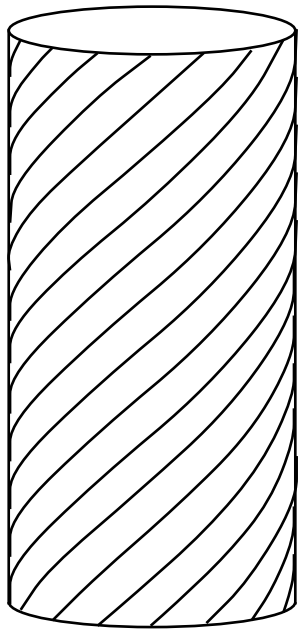

Figure 4

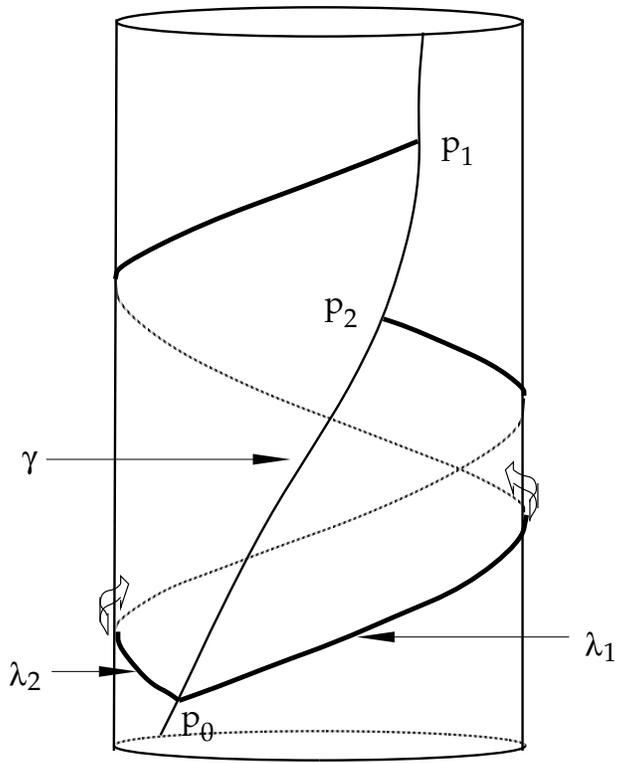

Figure 5